\documentclass[10pt]{iopart}
\usepackage{iopams}
\usepackage{graphicx,epsfig}% Include figure files
\usepackage{dcolumn}% Align table columns on decimal point
\usepackage{bm}% bold math

\def\fnl{f_{\mathrm{nl}}}

\newcommand{\Mpc}{{\mathrm{Mpc}}}
 
\begin{document}

%\preprint{APS:/123-QED}
\title{The Void Abundance with Non-Gaussian Primordial Perturbations}

\author{Marc Kamionkowski$^1$, Licia Verde$^{2}$, and Raul Jimenez$^{2}$}
{\it $^1$California Institute of Technology, Mail Code 130-33, Pasadena, CA 91125\\
$^2$ICREA \& Institute of Space Sciences (CSIC-IEEC), Campus UAB, Bellaterra, Spain}

\date{\today}

\begin{abstract}
We use a Press-Schechter-like calculation to study how the
abundance of voids changes in models with non-Gaussian initial
conditions.  While a positive skewness 
increases the cluster abundance, a negative skewness does the
same for the void abundance.  We determine the dependence of the
void abundance on the non-Gaussianity parameter $\fnl$ for the
local-model bispectrum---which approximates the bispectrum in
some multi-field inflation  models---and for the equilateral
bispectrum, which approximates the bispectrum in e.g. string-inspired DBI models of
inflation.  We show that the void abundance in
large-scale-structure surveys currently being considered should
probe values as small as $\fnl \lesssim 10$ and
$\fnl^{\mathrm{eq}}\lesssim 30$, over distance scales $\sim10$~Mpc.
\end{abstract}

%\pacs{}
\noindent{\it Keywords}: Cosmology: large scale structure of the Universe, superclusters and voids
\maketitle

\section{Introduction}
The paradigm of cosmological structure formation from a
spectrum of primordial perturbations like those predicted by
inflation has now been fairly well established by cosmic
microwave background (CMB) experiments \cite{cmbexperiments}.
We are thus now motivated to test more precisely
the predictions of inflation and to look for possible
deviations.  One of several such possibilities is measurement of
departures from Gaussianity of the initial
perturbations (see, e.g., Ref.~\cite{NGReview} and references therein). The
simplest slow-roll single-field models of inflation predict that
primordial perturbations should be very closely Gaussian
\cite{inflation}, but with predictably small departures from
Gaussianity \cite{localmodel,Equil}.  Multi-field \cite{larger}
models, such as the curvaton model \cite{curvaton}, and
string-inspired DBI \cite{Dvali:1998pa} inflationary
models can produce larger deviations from non-Gaussianity.

Departures from primordial Gaussianity can be sought in the
CMB \cite{Luo:1993xx,Verde:1999ij,Komatsu:2001rj},
large-scale structure (LSS) \cite{LSS}, and the
abundances and properties of the most massive
gravitationally-bound objects in the Universe today or at high
redshift \cite{abundances,MVJ00,Verde:2000vr}.  
The CMB provides a more powerful  and clean probe of primordial
non-Gaussianity than direct measurement of the bispectrum in
low-redshift LSS surveys in models with
scale-invariant non-Gaussianity \cite{Verde:1999ij}, although
biasing may amplify the effects of non-Gaussianity on LSS to the
level where they may be comparable in detectability to the CMB %LV
\cite{Dalal:2007cu, MV08, Slosar:2008hx,CVM08}.  Measurements of the cluster
abundance may do better than the CMB and LSS if the
non-Gaussianity is not scale-invariant \cite{LoVerde:2007ri}, as
may occur in DBI models.  What is clear, however, is
that the thorny systematic effects that enter in all of these
approaches will require that a variety of complementary avenues
be taken to establish a robust detection of non-Gaussianity. 

Voids have been considered as probes of cosmology, but no systematic study has been carried out for voids as probes of primordial non-gaussianity \cite{grossivoids}. In this paper, we consider the abundance of voids as a test of the distribution of the primordial perturbations.  Galaxy
clusters form at the highest 
overdensities of the primordial density field and thus probe
the high-density tail of the primordial density distribution
function.  Similarly, voids form in low-density regions and
should thus probe the low-density tail of the
distribution function.  If
there is a large negative skewness, the void-size distribution
function will be increased at the largest void sizes and
decreased at smaller void sizes, opposite to the effect on the
cluster mass function. 

In Section \ref{sec:PSabundance}, we develop a Press-Schechter
(PS) estimate of the void abundance for Gaussian initial
conditions.  This PS-like calculation is far from state of the art
\cite{Sheth:2003py,Furlanetto:2005cc} for Gaussian
initial conditions.  However, it is easily generalized to
non-Gaussian initial conditions and should be sufficiently
reliable to estimate the {\it fractional} effects of non-Gaussianity
on the void abundance (as it describes  well the halo abundance \cite{paper_in_prep}.  In Section
\ref{sec:skewness}, we discuss the relation between the skewness
and the non-Gaussian parameter $\fnl$ for the local model
\cite{localmodel}, which
approximates the non-Gaussianity in multi-field models, and
the equilateral model \cite{Equil}, which
approximates that in e.g. string-inspired DBI models.
In Section \ref{sec:results}, we provide results
of the void-abundance calculation, and we
estimate the smallest $\fnl$, for the local model and for the
equilateral model (with and without scale dependence) that
should be detectable in several surveys currently under study.
In Section \ref{sec:conclusions}, we make some concluding
remarks and outline further steps that must be taken before the
void abundance can be used to probe non-Gaussianity.

\section{The Press-Schechter Abundance}
\label{sec:PSabundance}

We begin by developing a PS--like calculation of
the void abundance, but we first review the standard
Press-Schechter calculation of the halo mass function.

Formation of a bound halo requires a linear-theory density
fluctuation $\delta_R>\delta_c\simeq1.69$, smoothed on scale
$R$.  The smoothing radius $R$ defines the halo mass $M$ via
$M=(4\pi/3) \rho_b R^3$, where $\rho_b$ is the mean
nonrelativistic-matter density.

The differential abundance of dark-matter halos  as a function
of mass and redshift is
\begin{equation}
     \frac{dn}{dM}=f\frac{\rho_b}{M}\left|
     \frac{dP(>\delta_c|z,M)}{dM}\right|,
\end{equation}
where $P(>\delta_c|z,M)$ denotes the probability that $\delta_R$
lie above the threshold for collapse $\delta_c$.  For Gaussian
initial conditions,
\begin{equation}
     P(>\delta_c| z, R) = \frac{1}{\sqrt{2\pi}}
     \frac{\sigma_R}{\delta_c}
     \exp\left[-\frac{1}{2}\frac{\delta_c^2}{\sigma_R^2}\right],
\end{equation}
where  $\sigma_R$ denotes the rms mass fluctuation on a  scale
$R$, and there is an implicit redshift dependence in
$\sigma_R$.  We then introduce the Press-Schechter
swindle---i.e., that every mass element in an underdense region
gets absorbed into the nearest overdensity---by introducing a
factor of 2 in the abundance.  The differential abundance of halos
is then
\begin{equation}
     \frac{dn}{dM} dM= \sqrt{ \frac{2}{\pi}} \frac{\rho_b}{M^2}
     \frac{\delta_c}{\sigma_M} \left| \frac{d \ln \sigma_M}{d
     \ln M} \right| e^{-\delta_c^2/2\sigma_M^2} dM.
\end{equation}
The mass function is then normalized so that $\int_0^\infty \, M
(dn/dM)\, dM=\rho_b$; i.e., every mass element in the Universe
is housed somewhere.

\subsection{Void abundance: Gaussian initial conditions}

A similar calculation can be used to estimate the void
abdundance.  In the PS description, every mass element
is in a gravitationally bound structure of some mass, and the
densities of these objects are all $\gtrsim200$ times the mean
density.  This means that $\gtrsim99.5\%$ of the volume of the
Universe is empty.  A void distribution function can be derived
in a way analogous to the PS mass function by realizing that
negative density fluctuations grow into voids (as opposed
to positive-density fluctuations, which grow into bound
objects), and that there is a critical underdensity $\delta_v$
for producing a void, to replace $\delta_c$, the critical
overdensity for producing bound objects.  In principle,
$\delta_v$ can be calculated from theory, but for now, we will
treat it as a phenomenological parameter.

The radius $R$ of a spherical volume in which a mass $M$ has
been cleared out is $R=(3\, M/4\pi \rho_b)^{1/3}$.  We thus
derive $(dn/dR)$, the differential abundance of voids of
diameter $R$,
\begin{equation}
     \frac{dn}{dR} = \frac{9}{2\pi^2} \sqrt{\frac{\pi}{2}} \frac{1}{R^4}
     \frac{|\delta_v|}{\sigma_M} \left| \frac{d \ln \sigma_M}{d
     \ln M} \right| e^{-\delta_v^2/2\sigma_M^2}.
\label{eqn:Gaussianabundance}
\end{equation}
We here introduce again the Press-Schechter swindle, to take
into account the fact that each underdense region expands by a
factor of two so that the entire (actually, 99.5\% of the)
volume of the Universe is occupied by a voids of radii $R$ with
the distribution $dn/dR$; i.e., the size distribution is
normalized so that
\begin{equation}
     \int_0^\infty\, \frac{dn}{dR} \frac{4\pi R^3}{3} \, dR =1.
\end{equation}

\subsection{Void abundance with non-Gaussian initial conditions}

The modification of Eq.~(\ref{eqn:Gaussianabundance}) when there
is a small primordial skewness $S_{3,M}$ (which may most
generally depend on the mass scale $M$) then follows from the
analogous modification for the halo abundance in
Refs.~\cite{MVJ00,Verde:2000vr,LoVerde:2007ri}.  The only
subtlety is that  $\delta_v$ is now a negative quantity (since
clusters come from overdensities, while voids come from
underdensities). In fact $P_{<\delta}=1-P_{>\delta}$ thus $|dP_{<\delta_v}/dM|=|dP_{>\delta_v}/dM|$.  
We can still use $|\delta_v|$ provided that
Eq.~(\ref{eqn:Gaussianabundance}) is
replaced by (see \cite{LoVerde:2007ri})
\begin{eqnarray}
     \frac{dn}{dR} &=& \frac{9}{2\pi^2} \sqrt{\frac{\pi}{2}}
     \frac{1}{R^4} e^{-\delta_v^2/2\sigma_M^2} 
     \left\{  \left| \frac{d \ln \sigma_M}{d \ln M}\right| 
     \left[   \frac{|\delta_v|}{\sigma_M} \right. \right. \nonumber \\
     & & \left. - \frac{S_3
     \sigma_M}{6} \left(\frac{\delta_v^4}{\sigma_M^4} -  2
     \frac{\delta_v^2}{\sigma_M^2} -1 \right) \right] \nonumber \\
     & & \left. + \frac{1}{6} \frac{dS_3}{dM} \sigma_M \left( \frac{
     \delta_v^2}{\sigma_M^2} -1 \right) \right\}.
\label{eqn:generalization}
\end{eqnarray}

\section{The skewness}
\label{sec:skewness}

 \begin{figure}
\begin{center}
\includegraphics[width=4.5in]{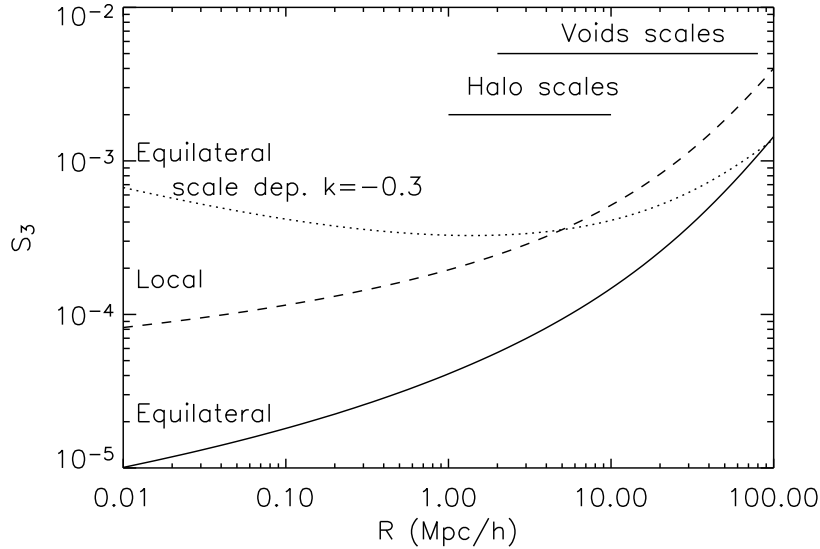}
\caption{Skewness $S_{3,R}$ for a non-Gaussianity parameter $\fnl=1$
     as function of radius $R$. The dashed line is the local non-Gaussian
     model; the solid line is the equilateral model; and the dotted line is
     the DBI-type equilateral model for a scale-dependence parameter
     $\kappa=-0.3$. See Ref.~\cite{LoVerde:2007ri} for details.
     We also indicate the scales probed by halo and void abundances.}
\label{fig:skew_R}
\end{center}
\end{figure}

According to Eq.~(\ref{eqn:generalization}), the void abundance
depends on the non-Gaussianity only through the skewness
$S_{3,R}$.  The skewness then depends on the detailed form of
the bispectrum.
There are a variety of bispectra considered in the literature.
The standard ``working-horse'' model for non-Gaussianity is
perhaps the local model, which features a bispectrum that arises
from multi-field inflation models.  There is then the
equilateral model, which approximates the bispectrum in DBI
models.  DBI models can also allow for a scale-dependent
non-Gaussianity, with power-law index $\kappa$.  See
Ref.~\cite{Equil,LoVerde:2007ri} for definitions and further details.

In Fig.~\ref{fig:skew_R}, we plot the skewness $S_{3,R}$ for a
non-Gaussianity parameter $\fnl=1$, as a function of 
$R$ for the local model, the equilateral model, and the
equilateral model with a scale-dependence parameter
$\kappa=-0.3$.  The skewness in the equilateral model is about a
factor 3 smaller than that for the local model. We thus infer
that constraints to $\fnl$ for the local model that derive
simply from $S_{3,R}$  will be a factor 3  more
stringent than for the equilateral model.  An equilateral model
with scale dependence characterized by $\kappa =-0.3$  will have
constraints similar to those for the local model.
 
We should also  note that $|\delta_v|$ is expected to be
smaller than $\delta_c$ by a factor $2-3$. If voids and halos
probe the same scales, then voids should yield  constraints on
$\fnl$ three times worse. However, voids may probe slightly
larger scales than halos, and as $S_3$ in these models
increases with scale, this compensates for $|\delta_v|$ being
smaller than $\delta_c$.
 
\section{Results}
\label{sec:results}

\begin{figure}
\begin{center}
\includegraphics[width=4.5in]{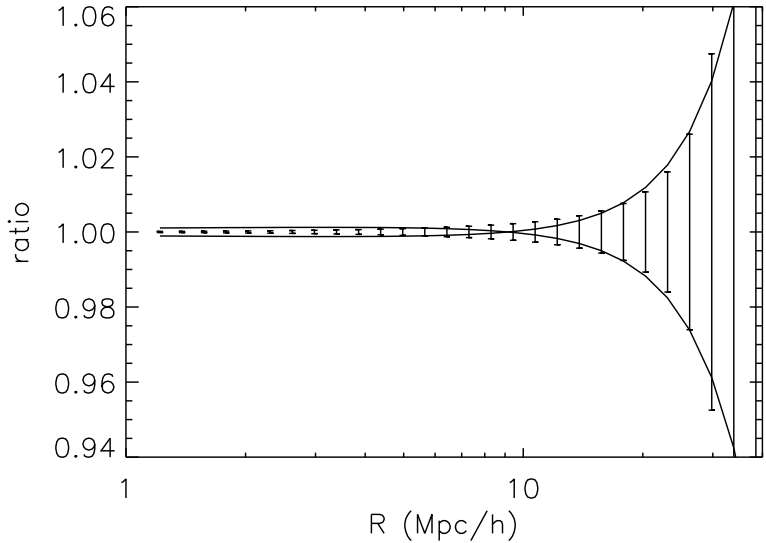}
\caption{The ratio of the void abundance with non-Gaussianity at
     the level of $|\fnl|=30$ to the void abundance with
     Gaussian initial conditions as a function of the void size
     $R$.  The curves are evaluated for a central redshift
     $z=0.8$, and the points are $1\sigma$ Poisson errors for a survey
     of width $\Delta z=0.3$ that covers
     30,000 square degrees for a fiducial Gaussian case.  The curve with a large ratio at
     large $R$ is for $\fnl<0$, while the curve with the small
     ratio at large $R$ is for $\fnl>0$.}
\label{fig:z08}
\end{center}
\end{figure}

To illustrate the promise held by the void abundance for
constraining $\fnl$, we consider several large-volume LSS
surveys at high redshift that are now being considered.  
The curves in Fig.~\ref{fig:z08} show the ratio of
the void abundance with non-Gaussianity to the abundance with
Gaussian initial conditions, as a function of void size $R$, for
representative values $\fnl=+30$ and $\fnl=-30$.  The curves are
evaluated for a redshift $z=0.8$.  The curve with greater void
abundance at larger $R$ and smaller void abundance at smaller
$R$ is for negative $\fnl$ and negative
skewness in the density field, while that with smaller (larger) void abundance at
larger (smaller) $R$ is for positive $\fnl$ (positive
skewness in the density field).  The crossover in $R$ between enhancement and
suppression of the void abundance reflects the onset
of the exponential tail of the PS distribution. 

The points with error bars estimate the $1\sigma$ error
bars (coming from Poisson errors in the void abundance)
anticipated for a survey with a redshift slice $\Delta
z=0.3$ centered at $z=0.8$ that covers 30,000 square degrees of
the sky.  The curves are evaluated for a critical underdensity
$\delta_v=-0.7$.  As indicated by the Figure, signal to noise
in the determination of $\fnl$ will come primarily from
the lower-$R$ end of the distribution, where the void abundance
is larger.

\begin{figure}
\begin{center}
\includegraphics[width=4.5in]{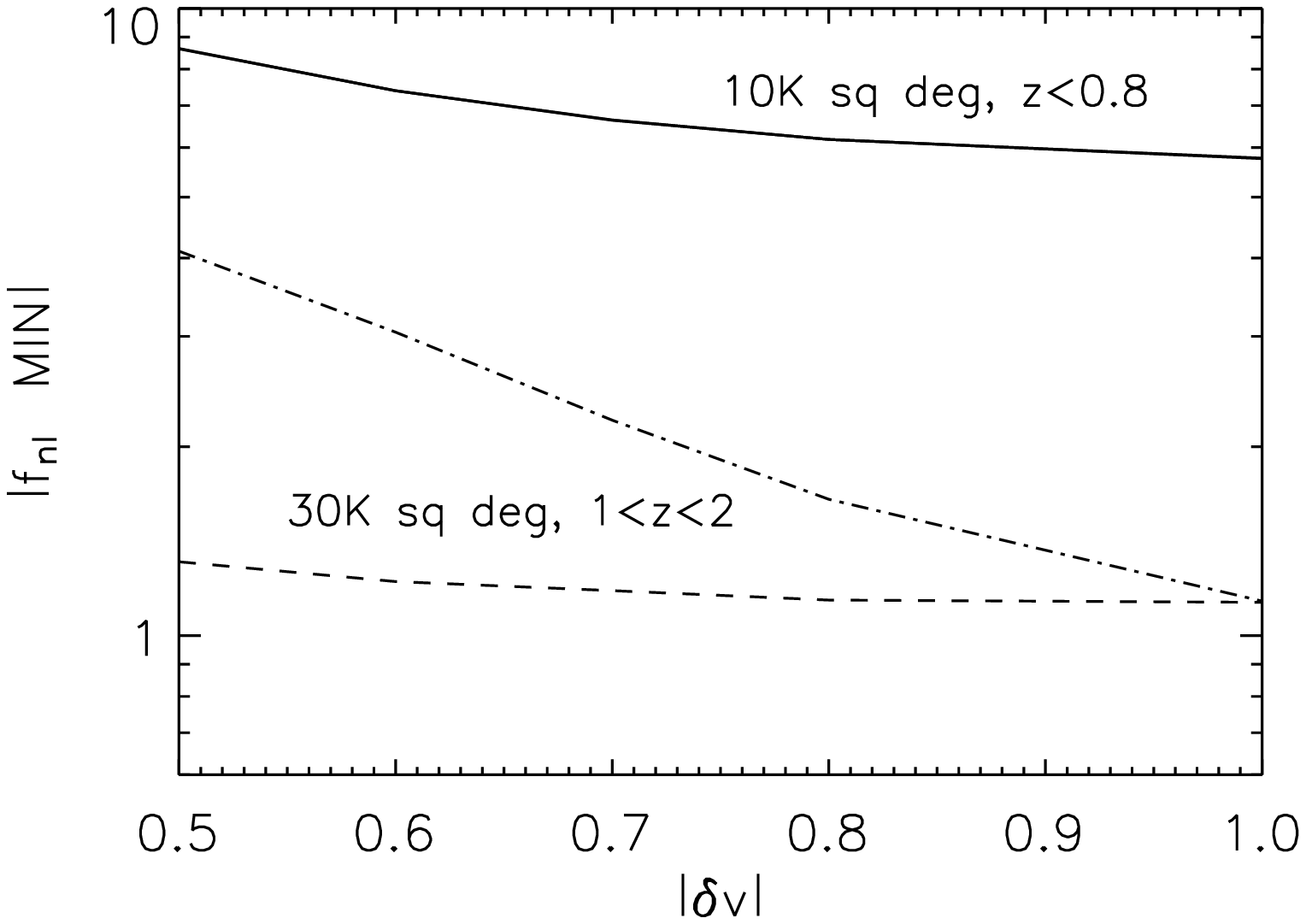}
\caption{The smallest local-model $\fnl$ detectable at the $1\sigma$ level,
     as a function of the critical underdensity $\delta_v$, for
     several surveys currently under study.  The solid 
     upper line is for something like the BOSS SDSS-3 survey
     using the void-size distribution over the range $2~h^{-1}~\Mpc<
     R <60~h^{-1}~\Mpc$.  The lower two (dashed)
     curves are for an ADEPT-like survey.  The upper dashed
     curve uses only the void distribution with
     $R>8\,h^{-1}~\Mpc$; the lower dashed curve uses
     $2~h^{-1}\,\Mpc> R< 60~h^{-1}\, \Mpc$.}
\label{fig:dvfnlmin}
\end{center}
\end{figure}

Fig.~\ref{fig:dvfnlmin} shows the smallest local-model $\fnl$ detectable at
the $1\sigma$ level as a function of the critical underdensity
$\delta_v$ for several surveys currently under study.  The solid
upper line is a survey with parameters comparable to those
proposed for the BOSS SDSS-3 survey \cite{boss} (which we parametrize as a
10,000 square-degree survey complete to redshift $z=0.8$); this
curve is evaluated using the void-size distribution over the
range $2~h^{-1}\, \Mpc < R < 60~h^{-1}\, \Mpc$.  The lower two (dashed)
lines assume parameters of a survey like ADEPT \cite{adept} (which we
parametrize as a 30,000 square-degree survey over the redshift
range $1<z<2$.  The upper dashed curve uses only the void
distribution with $R>8\,h^{-1}\,\Mpc$; the lower dashed curve
uses $2~h^{-1}\, \Mpc <R <60~h^{-1}\, \Mpc$.

Since the void abundance depends on non-Gaussianity only through
the dependence on $S_{3,R}$, we can use the curves shown in
Fig.~\ref{fig:skew_R} to compute constraints for  the equilateral model, both with and
without scale dependence.  As expected, since the equilateral
model (with no scale dependence) predicts a value of $S_{3,R}$
that is about 3 times smaller than in the local model, the smallest
the values of $\fnl^{\mathrm{eq}}$ accessible with the void
abundance is larger by roughly a factor of 3 than in the
local model.  Likewise, Fig.~\ref{fig:dvfnlmin} shows that
$\fnl$ for the equilateral model with $\kappa=-0.3$ is
comparable to that in the local model; the constraints to
$\fnl^{\mathrm{eq}}$ for $\kappa=-0.3$ are similar to
those in the local model.
This is particularly interesting as the large-scale bias effect due to  equilateral non-gaussianity  is  orders of magnitude smaller than the effect for the local case \cite{eq.biashalo}. Thus  the combination of the two measurements will  help discriminate among different type of non-gaussian initial conditions.

As Fig.~\ref{fig:dvfnlmin} shows, the results do not depend very
sensitively on the precise value of $\delta_v$.
The value that we choose for $\delta_v$ depends on the precise
definition of a void.  Here we have used as a canonical value
$\delta_v=-0.7$ for the critical linear-theory underdensity for
a void.  At this value of $\delta_v$, the physical void is
underdense by a factor of 2 (see, e.g., Fig. 1 in
Ref.~\cite{Furlanetto:2005cc}), which we believe to be a
conservative definition of a void.  If voids are defined in the
survey to be regions that are even emptier, then the relevant
magnitude $|\delta_v|$ will be even larger, and according to
Fig.~\ref{fig:dvfnlmin}, the smallest detectable $|\fnl|$ will
be even a bit smaller.

\section{Discussion}
\label{sec:conclusions}

The bottom line of our analysis is that the abundance of voids in
LSS surveys that are currently
being considered can probe values of the non-Gaussianity
parameter in the local model as small as $\fnl\sim10$ and in the
equilateral model (with no scale dependence) as small
$\fnl^{\mathrm{eq}}\sim 30$.  This
probe of non-Gaussianity may be competitive with
those coming from the CMB, LSS, and cluster
abundances \cite{Verde:1999ij,Verde:2000vr,LoVerde:2007ri}.  
They will complement CMB constraints by (a) providing a different
avenue with different systematic effects, and (b) probing
non-Gaussianity primarily over distance scales 2--60 Mpc,
scales generally smaller than those that will be probed by
the CMB.
The fact that voids constraints on  the different types of non gaussianity are comparable  is particularly interesting. In fact the large-scale bias effect due to  equilateral non-gaussianity  is  orders of magnitude smaller than the effect for the local case. Thus  the combination of the two measurements will  help discriminate among different type of non-gaussian initial conditions.

Finally, we note that we have here done no more than estimate
the smallest $\fnl$ detectable with the void abundance.  
To do so, we have taken a simple Press-Schechter approach to
estimate the fractional change in the void abundance.
While this approach should provide reasonable rough estimates to the
fractional change in the abundance, it will not reliably provide
the abundances themselves.  Before the void-abundance probe of
$\fnl$ can be implemented, we will therefore require
significantly more sophisticated calculations, which will probably
require numerical simulations, to accurately model not only the
void growth, but also the systematic effects that will arise in
any realistic void-identification algorithm.  We hope that
our results motivates this type of future work.

\section*{Acknowledgments}
MK thanks Andrew Benson for useful suggestions and the Aspen
Center for Physics for hospitality.  This work was supported at
Caltech by DoE DE-FG03-92-ER40701 and the
Gordon and Betty Moore Foundation. LV thanks M. LoVerde for
providing the curves of Fig. 3(a,b)  of Ref.~\cite{LoVerde:2007ri} in
table form. LV is supported by  FP7-PEOPLE-2007-4-3 IRG n 202182
and CSIC I3 grant 200750I034.  RJ is supported by grants from the
Spanish Science Ministry and The European Union (FP7). \\
%\end{acknowledgments}

\end{document}